\documentstyle[aps, preprint, epsf]{revtex}

\begin{document}
\title{Ward Identities for Interacting Electronic Systems}
\author{H. T. Nieh$^{1,2}$, Ping Sheng$^3$, and Xiao-Bing Wang$^1$}
\address{$^1$Center for Advanced Study, Tsinghua University, Beijing 100084, China\\
$^2$Institute for Theoretical Physics, State University of New York at\\
Stony Brook, Stony Brook, N.Y. 11794, U.S.A.\\
$^3$Department of Physics, The Hong University of Science and Technology,\\
Kowloon, Hong Kong}
\maketitle

\begin{abstract}
\indent A Ward-Takahashi identity, as a consequence of gauge invariance and
in a form that relates self-energy to the two-particle Bethe-Salpeter
scattering kernel, was first derived by Vollhardt and W\"{o}lfle for a
system of {\em independent} particles moving in a random medium. This is
generalized to a class of {\em interacting} electronic systems in materials
with or without random impurities, following a procedure previously used for
classical waves transport in disordered media. This class of systems also
possesses other symmetry properties such as invariance under time
translations and local spin rotations, which imply local conservation laws
for energy and spin current. They imply additional Vollhardt-W\"{o}lfle type
identities. We present non-perturbative derivations of these identities, and
consider the constraints they impose on the relationship between the
self-energy and the two-particle scattering kernel.
\end{abstract}


\newpage

Ward identities are statements of local conservation laws, and reflect the
corresponding symmetry properties of the system. The original Ward identity%
\cite{ward} and its generalization by Takahashi\cite{taka} were derived for
quantum electrodynamics, as a consequence of local current conservation, and
relate the current vertex function to the electron propagator. In
investigating the problem of the Anderson localization, Vollhardt and
W\"{o}lfle\cite{voll} considered a system of {\em independent} electrons
moving in a random potential, and derived a Ward-Takahashi identity in a
form that relates the electron self-energy to a two-particle Bethe-Salpeter
kernel induced by statistical randomness of the medium. More recently,
similar identities were derived for classical waves transport in disordered
media\cite{nieh}, for which, however, energy conservation underlies the
physics involved. It is clear from the latter derivations, which make use of
a Bethe-Salpeter type integral equation, that Vollhardt and W\"{o}lfle's
version of the Ward identity is valid for a considerably wider class of
systems, such as interacting electrons moving in materials with or without
random impurities. For this general class of electronic systems, energy and
spin, in addition to charge, are also conserved, reflecting invariance
properties of the system under time translation and spin rotation, and
implying additional Vollhardt-W\"{o}lfle type identities. We consider in
this paper the derivations of these identities, and their implications on
the self-energy.

The model Hamiltonian we consider is of the form,

\begin{eqnarray}
H &=&\int d^3x\left[ \frac 1{2m}\stackrel{\rightarrow }{\nabla }\overline{%
\Psi }(\vec{x},t)\cdot \stackrel{\rightarrow }{\nabla }\Psi (\vec{x},t)+V_1(%
\vec{x})\overline{\Psi }(\vec{x},t)\Psi (\vec{x},t)\right] \\
&&+\frac 12\int d^3xd^3x^{\prime }\overline{\Psi }(\vec{x},t)\Psi (\vec{x}%
,t)V_2(\vec{x}-\vec{x}^{\prime })\overline{\Psi }(\vec{x}^{\prime },t)\Psi (%
\vec{x}^{\prime },t),  \nonumber
\end{eqnarray}
where $\Psi $ is the two-component electron field, quantized according to
the usual equal-time anti-commutation relations, $V_1$ the potential
provided by atoms in the material, including randomly distributed static
impurities, and $V_2$ the interaction potential between any two electrons
with the property

\begin{equation}
V_2(\vec{x}-\vec{x}^{\prime })=V_2(\vec{x}^{\prime }-\vec{x}).
\end{equation}
The corresponding Lagrangian density is given by

\begin{eqnarray}
{\rm L} &=&i\overline{\Psi }(x)\stackrel{\cdot }{\Psi }(x)-\frac 1{2m}%
\stackrel{\rightarrow }{\nabla }\overline{\Psi }(x)\cdot \stackrel{%
\rightarrow }{\nabla }\Psi (x)-V_1(\vec{x})\overline{\Psi }(x)\Psi (x) \\
&&-\frac 12\int d^3x^{\prime }\overline{\Psi }(\vec{x},t)\Psi (\vec{x},t)V_2(%
\vec{x}-\vec{x}^{\prime })\overline{\Psi }(\vec{x}^{\prime },t)\Psi (\vec{x}%
^{\prime },t).  \nonumber
\end{eqnarray}
For convenience, we use the four-dimensional notations with $x^\mu =(t,x,y,z)
$ and $g^{\mu \nu }=(1,-1,-1,-1)$ for $\mu =0,1,2,3$. Due to the presence of
the non-local interaction term $V_2$ in the Lagrangian density, one should
exercise care in deriving the equation of motion. Instead of naively
applying the Euler-Lagrange equation, one considers the minimization of the
action, $\delta \int {\rm L}d^4x$, to derive the expected equation of motion
:

\begin{equation}
i\stackrel{\cdot }{\Psi }=\left( -\frac 1{2m}\nabla ^2+V_1\right) \Psi +\int
d^3x^{\prime }\overline{\Psi }(\vec{x}^{\prime },t)\Psi (\vec{x}^{\prime
},t)V_2(\vec{x}-\vec{x}^{\prime })\Psi (x).
\end{equation}
The Lagrangian density is invariant under the gauge transformation,

\begin{eqnarray}
\Psi (x) &\longrightarrow &e^{i\lambda }\Psi (x), \\
\overline{\Psi }(x) &\longrightarrow &e^{-i\lambda }\overline{\Psi }(x), 
\nonumber
\end{eqnarray}
the time translation,

\begin{eqnarray}
t &\longrightarrow &t^{\prime }=t+\varepsilon , \\
\Psi (\vec{x},t) &\longrightarrow &\Psi (\vec{x},t^{\prime }-\varepsilon ), 
\nonumber \\
\overline{\Psi }(\vec{x},t) &\longrightarrow &\overline{\Psi }(\vec{x}%
,t^{\prime }-\varepsilon ),  \nonumber
\end{eqnarray}
and the spin rotation

\begin{eqnarray}
\Psi (x) &\longrightarrow &e^{i\vec{\omega}\cdot \frac{\vec{\sigma}}2}\Psi
(x), \\
\overline{\Psi }(x) &\longrightarrow &\overline{\Psi }(x)e^{-i\vec{\omega}%
\cdot \frac{\vec{\sigma}}2},  \nonumber
\end{eqnarray}
where $\vec{\sigma}$ are the Pauli spin matrices. The corresponding
conserved quantities are charge, energy, and spin, respectively.

In disordered media, there is a random distribution of impurities. We need
to consider a statistical ensemble of impurity configurations and average
over the ensemble. Under rather general assumptions on the statistical
ensemble, as has been fully discussed by, e.g., Frisch\cite{fris},
statistical averaging results in an effective electron-electron interaction,
which adds to the real electron-electron interaction potential $V_2$.

Local current conservation as a consequence of gauge invariance is of the
form

\begin{equation}
\partial _\mu j^\mu =0,
\end{equation}
where

\begin{equation}
j^0=\overline{\Psi }\Psi ,\text{ \qquad }j^k=\frac i{2m}[\overline{\Psi }%
\partial ^k\Psi -(\partial ^k\overline{\Psi })\Psi ].
\end{equation}
It implies that

\begin{eqnarray}
&&\partial _\mu ^x\langle 0|T\{j^\mu (x)\Psi (y)\overline{\Psi }%
(z)\}|0\rangle \\
&=&-\delta ^4(x-y)\langle 0|T\{\Psi (x)\overline{\Psi }(z)\}|0\rangle
+\delta ^4(x-z)\langle 0|T\{\Psi (y)\overline{\Psi }(x)\}|0\rangle  \nonumber
\\
&=&-\delta ^4(x-y)\frac 1iG(x,z)+\delta ^4(x-z)\frac 1iG(y,x).  \nonumber
\end{eqnarray}
The above equation holds for any single configuration of the impurities.
When statistical averaging on an ensemble of impurity configurations is
applied to it, an additional statistical interaction infuses itself into the
structure of the expressions on both sides of the equation. An irreducible
current vertex function is defined by

\begin{equation}
\langle 0|T\{j_\mu (x)\Psi (y)\overline{\Psi }(z)\}|0\rangle ^{av}=\int
d^4yd^4z\frac 1iG^{(e)}(y-\eta )\Gamma _\mu (\eta |x|\zeta )\frac 1i%
G^{(e)}(\zeta -z),
\end{equation}
where expressions on both sides of the equation refer to configurationally
averaged quantities, and use has been made of the fact that averaging
restores translational invariance and, as a consequence, the averaged
propagation function, $G^{(e)}$, depends only on the difference of the two
coordinates. Averaging applied to both sides of (10) leads to

\begin{eqnarray}
&&\partial _x^\mu \int d^4\eta d^4\zeta \frac 1iG^{(e)}(y-\eta )\Gamma _\mu
(\eta |x|\zeta )\frac 1iG^{(e)}(\zeta -z) \\
&=&-\delta ^4(x-y)\frac 1iG^{(e)}(x-z)+\delta ^4(x-z)\frac 1iG^{(e)}(y-x), 
\nonumber
\end{eqnarray}
which implies the Ward-Takahashi identity :

\begin{equation}
\partial _x^\mu \Gamma _\mu (\eta |x|\zeta )=G^{(e)-1}(\eta -x)\frac 1i%
\delta ^4(x-\zeta )-\delta ^4(\eta -x)\frac 1iG^{(e)-1}(x-\zeta ).
\end{equation}

For a system of non-interacting electrons moving in disordered media,
Vollhardt and W\"{o}lfle\cite{voll}, using diagram considerations, derived
their version of the Ward-Takahashi identity that relates the electron
self-energy with the two-particle Bethe-Salpeter kernel instead of the
current vertex function. Rather than following their analysis in terms of
diagrams, we here adopt the non-perturbative procedure previously used for
the derivation of energy-conservation Ward-Takahashi identities for
classical waves transport in disordered media\cite{nieh}.

A crucial input in the non-perturbative derivation is the following integral
equation for the vertex function $\Gamma _\mu $,

\begin{eqnarray}
&&\Gamma _\mu (\eta |x|\zeta )-\Gamma _\mu ^{(0)}(\eta |x|\zeta ) \\
&=&\int d^4yd^4y^{\prime }d^4zd^4z^{\prime }\frac 1iG^{(e)}(y^{\prime
}-y)\Gamma _\mu (y|x|z)\frac 1iG^{(e)}(z-z^{\prime })K(\eta \zeta ,y^{\prime
}z^{\prime }),  \nonumber
\end{eqnarray}
where $K$ is the Bethe-Salpeter kernel, and $\Gamma _\mu ^{(0)}$ denotes the
vertex function corresponding to a reference homogeneous medium and in the
absence of the interaction $V_2$. It is of identical form as the one in
quantum electrodynamics\cite{bjor}. The validity of this integral equation,
schematically illustrated in Fig. 1, is on the same footing as the
Bethe-Salpeter equation. We note that all quantities in the above integral
equation refer to statistically averaged ones. Combining the two equations
(13) and (14) results in the Ward identity in the Vollhardt-W\"{o}lfle form :

\begin{eqnarray}
&&\Sigma (\eta -x)\delta ^4(x-\zeta )-\Sigma (x-\zeta )\delta ^4(\eta -x) \\
&=&\int d^4yd^4z[\delta ^4(y-x)G^{(e)}(x-z)-\delta
^4(x-z)G^{(e)}(y-x)]U(z-\zeta ,\eta -y,\vec{y}-\vec{z}),  \nonumber
\end{eqnarray}
where the self-energy $\Sigma $ is defined by

\begin{equation}
G^{(e)-1}=G^{(0)-1}-\Sigma
\end{equation}
and $U$ is an alternative expression for $K$, with specified coordinate
dependence. The dependence of $K$ on the coordinates is inferred from the
following two properties. First, because of the restored translational
invariance due to averaging, only three of the four coordinates in $K$ are
independent, and are chosen to be $z-\zeta ,\eta -y,$ and $y-z$. Second,
since the two channels are mediated by the effective statistical interaction
and $V_2$, both of which are static, ($y-z$) can actually be replaced by ($%
\vec{y}-\vec{z}$). In momentum space, the identity takes the form

\begin{equation}
\Sigma (q+p)-\Sigma (q)=\int \frac{d^3\omega }{(2\pi )^3}[G^{(e)}(q+\vec{%
\omega})-G^{(e)}(q+p+\vec{\omega})]U(q,q+p,-\vec{\omega}).
\end{equation}
This is the Ward identity derived by Vollhardt and W\"{o}lfle\cite{voll},
but here extended to include the non-local interaction $V_2$ between any two
electrons.

The Lagrangian density is invariant under the time translation. This implies
the local conservation law

\begin{equation}
\partial ^\mu T_{\mu 0}=0,
\end{equation}
where

\begin{eqnarray}
T_{00}(x) &=&\frac 1{2m}\stackrel{\rightarrow }{\nabla }\overline{\Psi }%
(x)\cdot \stackrel{\rightarrow }{\nabla }\Psi (x)+V_1(\vec{x})\overline{\Psi 
}(x)\Psi (x)  \nonumber \\
&&+\frac 12\int d^3x^{\prime }\overline{\Psi }(\vec{x}^{\prime },t)\Psi (%
\vec{x}^{\prime },t)V_2(\vec{x}-\vec{x}^{\prime })\overline{\Psi }(x)\Psi
(x), \\
T_{k0}(x) &=&\frac 1{2m}[\partial _k\overline{\Psi }(x)\partial _0\Psi
(x)+\partial _0\overline{\Psi }(x)\partial _k\Psi (x)].  \nonumber
\end{eqnarray}
In particular, the energy density is $T_{00}$. Similar to the case of
current conservation, we consider in the present case the divergence of the
three-point function $\langle 0|T\{T_{\mu 0}(x)\Psi (y)\overline{\Psi }%
(z)\}|0\rangle $. To proceed, we need to know the equal-time commutation
relations $[T_{00}(x),\Psi (y)]$, etc.. Due to the presence of the non-local
interaction $V_2$, we consider the commutation relation with the integrated
total energy:

\begin{eqnarray}
&&[\int d^3xT_{00}(\vec{x},t),\Psi (\vec{y},t)] \\
&=&-\int d^3x\delta ^3(\vec{x}-\vec{y})\left\{ \left[ -\frac 1{2m}\nabla
^2+V_1(\vec{x})\right] \Psi +\int d^3x^{\prime }\overline{\Psi }(\vec{x}%
^{\prime },t)\Psi (\vec{x}^{\prime },t)V_2(\vec{x}-\vec{x}^{\prime })\Psi
(x)\right\}  \nonumber \\
&&-\frac 1{2m}\int d^3x\stackrel{\rightarrow }{\nabla }\cdot [\delta ^3(\vec{%
x}-\vec{y})\stackrel{\rightarrow }{\nabla \Psi (x)}]  \nonumber
\end{eqnarray}
There is some uncertainty as to what to do with the 3-dim divergence term in
the above equation. We have considered various arguments for or against its
inclusion in the local equal-time commutation relation. The most convincing
argument against its inclusion comes from an exercise in deriving the Ward
identities by using the path-integral formalism. We thus conclude that, on
account of (4),

\begin{equation}
\lbrack T_{00}(\vec{x},t),\Psi (\vec{y},t)]=-\delta ^3(\vec{x}-\vec{y})%
\stackrel{\cdot }{\Psi }(\vec{x},t),
\end{equation}
and similarly for $\overline{\Psi }$. Consequently,

\begin{equation}
\partial _x^\mu \langle 0|T\{T_{\mu 0}(x)\Psi (y)\overline{\Psi }%
(z)\}|0\rangle =-\delta ^4(x-y)\frac \partial {\partial t_x}G(x,z)-\delta
^4(x-z)\frac \partial {\partial t_x}G(y,x).
\end{equation}
Defining the energy vertex function $\Gamma _\mu ^{(E)}$ according to

\begin{equation}
\langle 0|T\{T_{\mu 0}(x)\Psi (y)\overline{\Psi }(z)\}|0\rangle ^{av}=\int
d^4\eta d^4\zeta \frac 1iG^{(e)}(y-\eta )\Gamma _\mu ^{(E)}(\eta |x|\zeta )%
\frac 1iG^{(e)}(\zeta -z)
\end{equation}
we obtain from (22) the relation

\begin{equation}
\partial _x^\mu \Gamma _\mu ^{(E)}(\eta |x|\zeta )=G^{(e)-1}(\eta -x)\frac %
\partial {\partial t_x}\delta ^4(x-\zeta )+G^{(e)-1}(x-\zeta )\frac \partial
{\partial t_x}\delta ^4(\eta -x).
\end{equation}
which is the Ward-Takahashi identity for energy conservation. Compared with
the current-conservation Ward-Takahashi identity (13), there are here in
(24) extra linear time derivatives, which are the unmistakable signatures
for energy. The two identities describe charge flow and energy flow,
respectively.

To derive the corresponding Vollhardt-W\"{o}lfle type identity, we again
make use of an integral equation for the energy vertex function, which is
identical in structure to the charge counterpart,

\begin{eqnarray}
&&\Gamma _\mu ^{(E)}(\eta |x|\zeta )-\Gamma _\mu ^{(E)(0)}(\eta |x|\zeta ) 
\nonumber \\
&=&\int d^4yd^4y^{\prime }d^4zd^4z^{\prime }\frac 1iG^{(e)}(y^{\prime
}-y)\Gamma _\mu ^{(E)}(y|x|z)\frac 1iG^{(e)}(z-z^{\prime })K(\eta \zeta
;y^{\prime }z^{\prime }).
\end{eqnarray}
Combining (24) and (25), we obtain the desired identity

\begin{eqnarray}
&&\Sigma (\eta -x)\frac \partial {\partial t_x}\delta ^4(x-\zeta )+\Sigma
(x-\zeta )\frac \partial {\partial t_x}\delta ^4(\eta -x) \\
&=&\int d^4yd^4z[\delta ^4(y-x)\frac \partial {\partial t_x}%
G^{(e)}(x-z)+\delta ^4(x-z)\frac \partial {\partial t_x}G^{(e)}(y-x)]U(z-%
\zeta ,\eta -y,\vec{y}-\vec{z})  \nonumber
\end{eqnarray}
In momentum space, it is of the form

\begin{eqnarray}
&&q_0\Sigma (q+p)-(q_0+p_0)\Sigma (q)  \nonumber \\
&=&\int \frac{d^3\omega }{(2\pi )^3}[q_0G^{(e)}(q+\vec{\omega}%
)-(q_0+p_0)G^{(e)}(q+p+\vec{\omega})]U(q,q+p,-\vec{\omega}).
\end{eqnarray}

The Lagrangian density (3) is also invariant under local spin rotations, for
which

\begin{equation}
\Psi (x)\longrightarrow e^{i\vec{\omega}\cdot \frac{\vec{\sigma}}2}\Psi (x),%
\text{ \quad }\overline{\Psi }(x)\longrightarrow \overline{\Psi }(x)e^{-i%
\vec{\omega}\cdot \frac{\vec{\sigma}}2}
\end{equation}
This symmetry property implies the local conservation law

\begin{equation}
\partial _\mu S_a^\mu =0,\text{ \qquad }a=1,2,3,
\end{equation}
where the spin current is given by

\begin{eqnarray}
S_a^0 &=&\overline{\Psi }\frac{\sigma _a}2\Psi , \\
\stackrel{\rightarrow }{S_a} &=&\frac i{2m}(\overline{\Psi }\frac{\sigma _a}2%
\stackrel{\rightarrow }{\nabla }\Psi -\stackrel{\rightarrow }{\nabla }%
\overline{\Psi }\frac{\sigma _a}2\Psi ).  \nonumber
\end{eqnarray}
Using a procedure in complete parallel to the cases of gauge and
time-translation symmetries, we obtain the following Vollhardt-W\"{o}lfle
type identity for $a=1,2,3,$

\begin{eqnarray}
&&\Sigma (\eta -x)\frac{\sigma _a}2\delta ^4(x-\zeta )-\delta ^4(\eta -x)%
\frac{\sigma _a}2\Sigma (x-\zeta ) \\
&=&\int d^4yd^4z[\delta ^4(y-x)\frac{\sigma _a}2G^{(e)}(x-z)-\delta
^4(x-z)G^{(e)}(y-x)\frac{\sigma _a}2]U(z-\zeta ,\eta -y,\vec{y}-\vec{z}) 
\nonumber
\end{eqnarray}
In momentum-space representation, it is of the form

\begin{equation}
\Sigma (q+p)\frac{\sigma _a}2-\frac{\sigma _a}2\Sigma (q)=\int \frac{%
d^3\omega }{(2\pi )^3}[\frac{\sigma _a}2G^{(e)}(q+\vec{\omega})-G^{(e)}(q+p+%
\vec{\omega})\frac{\sigma _a}2]U(q,q+p,-\vec{\omega}).
\end{equation}

The collection of the three Vollhardt-W\"{o}lfle identities (15), (26), and
(31), or their momentum-space counterparts (17), (27), and (32) set fairly
stringent conditions on the relationship between the self-energy and the
two-particle scattering kernel. Combining (17) and (27) yields the
surprisingly simple but strong relation

\begin{equation}
\Sigma (q)=\int \frac{d^3\omega }{(2\pi )^3}G^{(e)}(q+p+\vec{\omega}%
)U(q,q+p,-\vec{\omega}).
\end{equation}
In configuration space, it is of the form

\begin{equation}
\delta ^4(\eta -x)\Sigma {}(\eta -\zeta )=\int d^4yd^4z\delta
^4(x-z)G^{(e)}(y-z)K(\eta \zeta ;yz).
\end{equation}
We have checked and verified this relation in terms of low-order diagrams.
According to (33) or (34), all self-energy diagrams can be generated from
the corresponding irreducible kernel diagrams. For example, the self-energy
diagram in Fig.2a can be generated from the irreducible kernel diagram in
Fig.2b by connecting with a propagator as shown in Fig.2c. In the
verification of the relation (34), we have disposed of the tadpole type
self-energy diagrams, such as in Fig.3a, which can not be generated
according to (34). The disposal of these tadpole type diagrams is based on
the following considerations. First, in the traditional formulation\cite
{itzy} of the Lagrangian field theory, the Lagrangian density is defined in
terms of normal ordered products of the field operators. All tadpole
diagrams, by the very definition of the normal ordering, do not appear at all%
\cite{itzy}. Second, if for any reason, normal ordering of the products is
not adopted, as is effectively the case in the path-integral formulation of
quantum field theory, tadpole diagrams do appear. However, their
contribution to the self-energy is ``eliminated'' order by order by the
``mass'' insertion diagram in Fig.3b, due to the presence of the mass
counter term, actually the chemical-potential counter term in the present
context, which is introduced in the Lagrangian density to achieve
renormalization. We note that renormalization counter terms are introduced
in either of the above two formulations, although details of their functions
differ somewhat. For example, in the case on hand, the ``mass'' counter term
absorbs the (infinite) tadpole contribution to the self-energy in the
path-integral approach, while in the traditional approach of Lagrangian
field theory, because the tadpole diagrams do not appear, there is no need
for the ``mass'' counter term to absorb them.

The strong relationship (34) perhaps reflects the fact that both the charge
and the energy are exclusively carried by the same carriers, the electrons.
It would be of interest to investigate additional implications of the set of
three Vollhardt-W\"{o}lfle identities discussed in this note.

\acknowledgments

One of authors (HTN) would like to thank Dr. Guang-Ming Zhang for helpful
discussions and a question that led to the consideration of the
spin-rotation symmetry, and another (PS) acknowledges the support of Hong
Kong Research Grants Council Grant No. HKUST685/96P for this work.

\begin{figure}
\epsfxsize=0.8\columnwidth
  \begin{center}
\epsffile{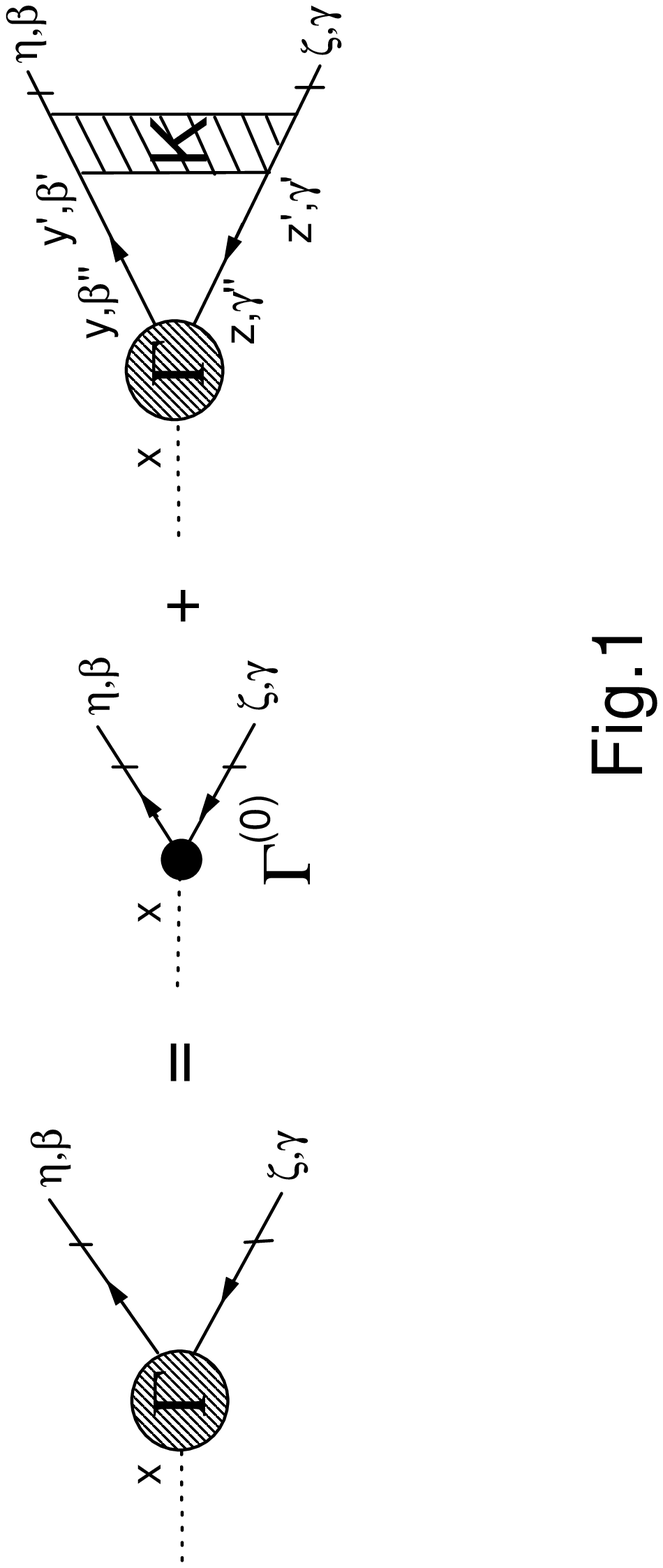}
  \end{center}
\caption{Schematic content of Eq. (14) and (25).}
\end{figure}

\begin{figure}
\epsfxsize=0.8\columnwidth
  \begin{center}
\epsffile{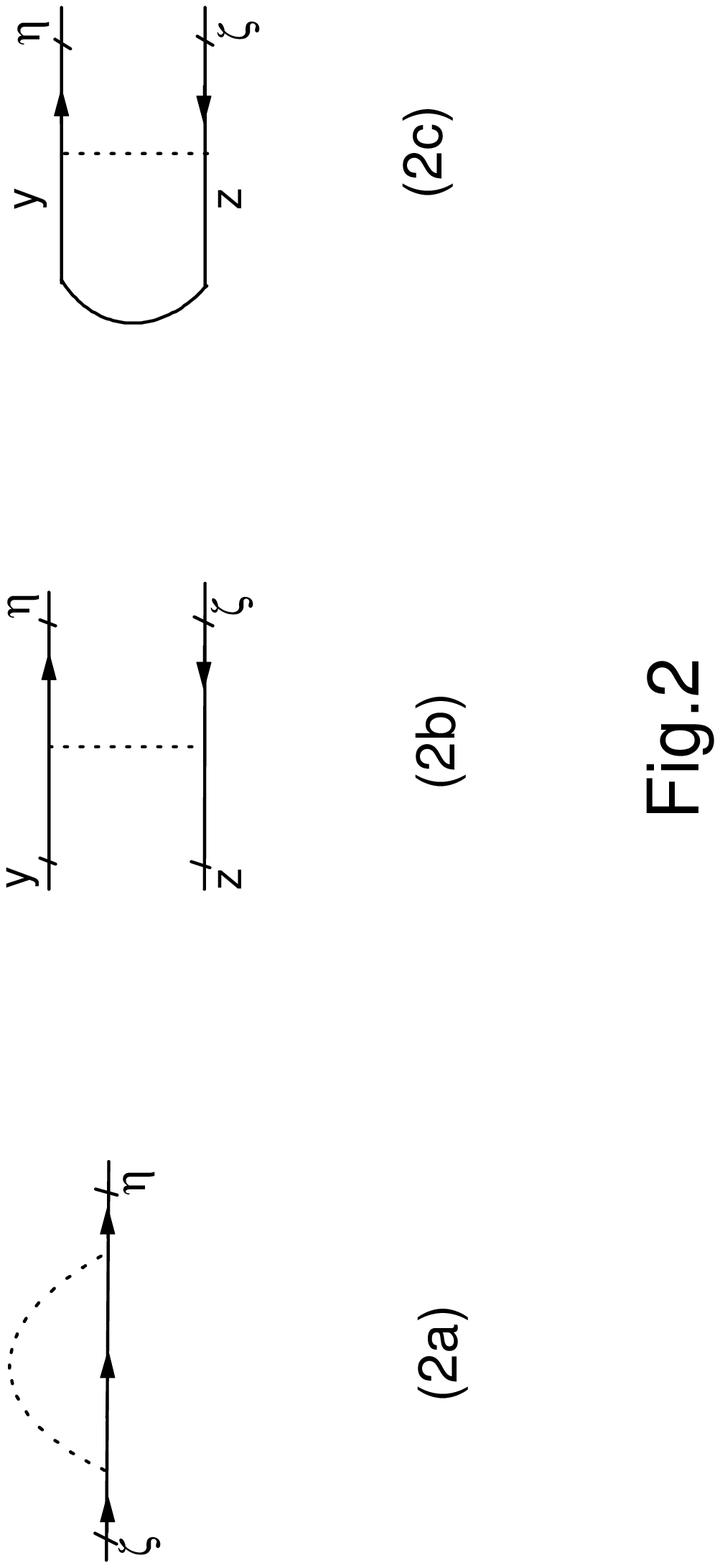}
  \end{center}
\caption{Schematic relationship between $\Sigma $ and $K$.}
\end{figure}

\begin{figure}
\epsfxsize=0.8\columnwidth
  \begin{center}
\epsffile{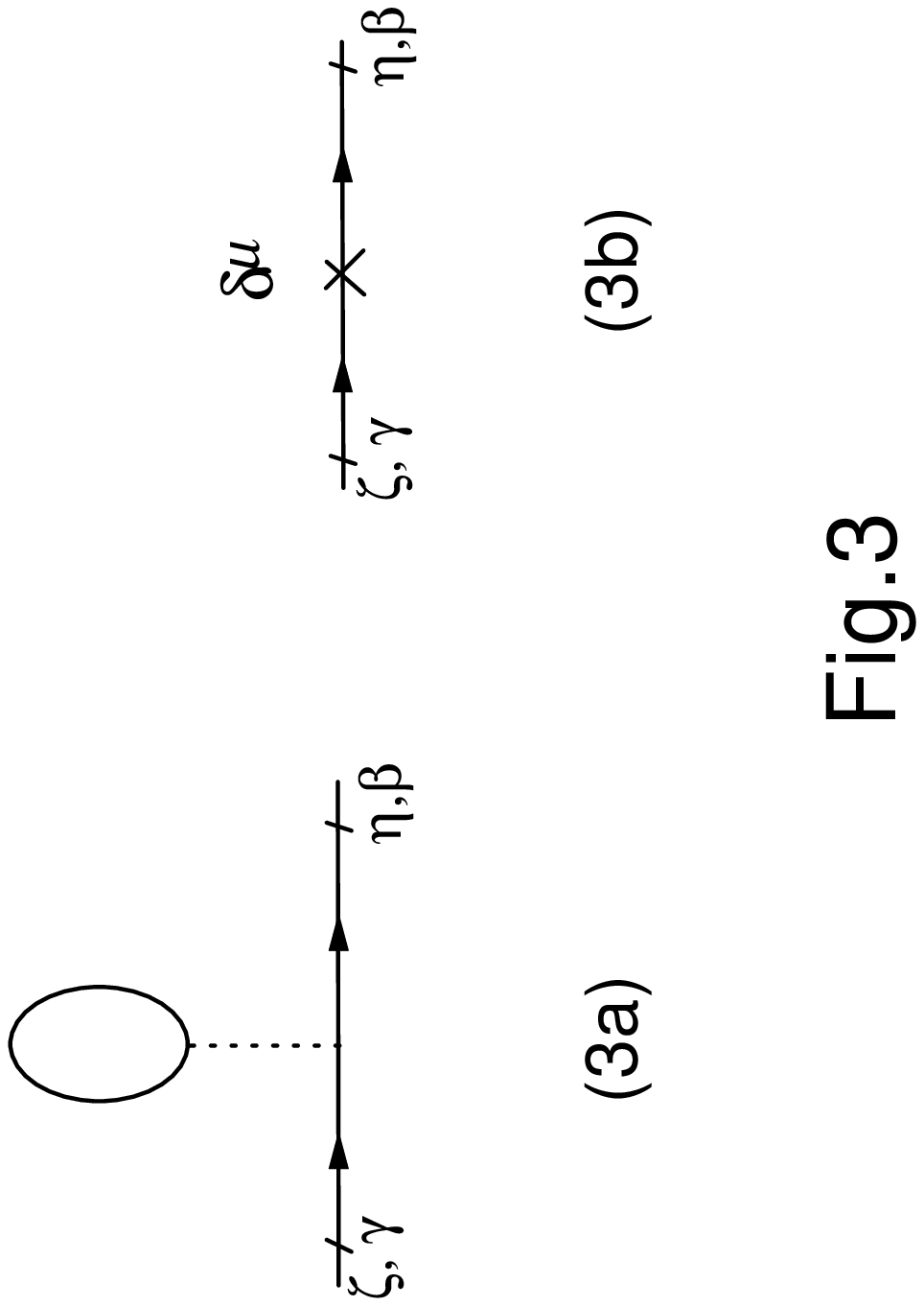}
  \end{center}
\caption{``Tadpole'' self-energy.}
\end{figure}

\end{document}